# Refractometry with filled antiresonant capillary fibers

Sinei Felix Gonçalves Junior [1], Lidia O. Rosa [2], Flavio A. M. Marques [1], Alexandre A. C. Cotta [1], Jefferson E. Tsuchida [1], Eric Fujiwara [3], Cristiano M. B. Cordeiro [2], Jonas H. Osório [1,*]

[1] *Multi-user Optics and Photonics Laboratory, LaMOF, Department of Physics, Federal University of Lavras, Lavras, 37200-900, Brazil*
[2] *Institute of Physics Gleb Wataghin, University of Campinas, Campinas, 13083-859, Brazil*
[3] *School of Mechanical Engineering, University of Campinas, Campinas, 13083-860, Brazil*

*\*jonas.osorio@ufla.br*

**We demonstrate the realization of refractometric measurements relying on the study of the transmission spectrum of filled capillary fibers. In this method, the fiber is filled with a material with a lower refractive index than that of the capillary and, due to antiresonant guidance mechanism, a characteristic transmission spectrum alternating high and low attenuation regions is obtained. The refractive index data is hence extracted by analyzing the spectral positions of the fiber transmission bands. While this method holds broad applicability for diverse materials, we specifically applied this technique to characterize agarose gels, due to their interest as a promising optical material. By analyzing the transmission spectra across the 600-900 nm wavelength range, we determined the dispersion trend for agar gels prepared with varying water and glycerol concentrations and estimated their first-order Sellmeier coefficients. The reported refractometric method provides a simple and promising means for characterizing the dispersion properties of a wide range of materials, including gels, solids, and liquids, also opening new possibilities for the development of new refractive index sensing platforms.**

Refractometric measurements play a fundamental role in a wide array of scientific and industrial applications, from material characterization to chemical sensing and biomedical diagnostics. In this context, information on a material refractive index provides valuable insights into its composition, density, and molecular interactions. For instance, it is an important parameter in determining the concentration of solutions and characterizing materials for photonics and optoelectronics applications [1-3].

However, typical refractometric techniques, such as those based on Abbe refractometers [4] or prism-coupling methods [5], often require the use of specialized and dedicated equipment. Indeed, these setups can be cumbersome and demand specific sample preparation methods. Furthermore, many standard methods rely on measurements at single wavelengths, or a limited number of discrete wavelengths, making the determination of material dispersion a complex and time-consuming process. Thus, there is a need for the development of more versatile solutions capable of providing data on the dispersion of materials.

In another framework, hollow-core optical fibers (HCF), including capillaries, have emerged as promising platforms for sensing applications due to their ability to guide light within an air or liquid core [6, 7]. Silica capillaries can guide visible and near-infrared light through the antiresonant reflecting optical waveguide mechanism (ARROW), in which light guidance arises from interference processes due to reflections on the inner and outer walls of the capillary [8]. This principle makes these simple structures viable waveguides for optical measurements and applications.

Here, we present a simple configuration that explores the transmission properties of filled silica capillary fibers for attaining refractometric measurements. By analyzing the transmission spectra of capillaries filled with materials with refractive indices (lower than that of silica), we could retrieve the refractive index of the samples across the 600-900 nm wavelength range. Whilst this method is applicable to diverse materials, for this study, we applied this technique to characterize agarose gels, owing to their interest as promising optical material. Agarose is a promising optical material due to its biocompatibility, transparency, and the ability to have its properties tuned by adjusting its concentration and composition. Additionally, it can act as a versatile medium for encapsulating and characterizing substances of interest [9].

Remarkably, this configuration can directly measure the absolute values of the refractive index of the samples. This stands in contrast to many existing fiber-optic refractive index sensors, which, while offering high sensitivity and compactness, typically provide information only on refractive index variations relative to a known reference [1-3]. Thus, such sensors often lack the capability to yield the absolute refractive index value, hence limiting their utility for material characterization. While previous studies have characterized the dispersion of liquids using more complex HCFs [7], our approach is significantly simpler as it relies on silica capillary tubes. Therefore, the research reported herein demonstrates a straightforward method for characterizing the dispersion of materials while holding promise for broader applications in characterizing liquids and solids, as well as advancing refractive index sensing technologies.

The platform investigated in this study stands for a silica capillary tube with a wall thickness $t$ and refractive index $n_{tube}$. In the demonstrations reported herein, the capillary is filled with an agar gel possessing a refractive index $n_{agar}$. The typical transmission spectrum of such a filled silica capillary displays alternating regions of high and low attenuation. The wavelengths corresponding to the transmission minima, $\lambda_m$, can be described by Eq. (1), where $m$ is an integer [10].

Information on the spectral position of the transmission minima can, therefore, allow for estimating $n_{agar}$ if $t$, $n_{tube}$, and $m$ are known. In the analyses to be reported in the following, we set $n_{tube}$ as the refractive index of silica considering its Sellmeier expression [11].

$$\lambda_m = \frac{2t}{m}\sqrt{n_{tube}^2 - n_{agar}^2} \quad (1)$$

The preparation of the agar-core silica-cladding fiber involves a straightforward process, as depicted in Fig. 1. Initially, the agar solution is prepared by dissolving the desired concentration of the food-grade agar powder in distilled water (or a water-glycerol mixture) under controlled heating on a hot plate until a homogeneous solution is achieved (Fig. 1a). Following this, a second step is carried out, involving the removal of entrapped air bubbles, which can adversely affect the homogeneity of the gel. This is achieved by allowing the solution to cool and solidify, and by carefully cutting away and discarding portions of the gel where bubble formation has been observed (Fig. 1b). In sequence, an empty silica capillary fiber is gently inserted into the solidified agar (Fig. 1c) so that the capillary is completely filled by the gel. Finally, the filled capillary is removed from the agar, resulting in the desired agar-filled silica capillary fiber (Fig. 1d). Here, we used silica capillaries displaying a wall thickness of 140 µm, a core diameter of 2.55 mm, and a length of 7 cm.

Fig. 1e illustrates the experimental setup employed for measuring the transmission spectrum of the agar-filled silica capillary fibers. The optical beam emitted from a supercontinuum source (SC, NKT SuperK COMPACT) is focused by a first lens (L1, 5× objective lens) onto the input facet of the agar-filled capillary fiber. After propagating through the fiber, the transmitted light is collected by a second lens (L2, 5× objective lens) and directed towards a spectrometer (SPTR, Ocean Optics). A third lens (L3, 20× objective lens) is used to focus the light onto the spectrometer collecting fiber. Additionally, we used an iris (I) to minimize the collection of residual light traveling through the capillary wall and a neutral density filter (ND) to prevent detector saturation. This configuration allowed for the acquisition of the transmission spectra, from which the agar refractive index can be derived. The measurements reported in this paper have been performed at room temperature.

Fig. 2a displays the normalized transmission spectra (maximum transmission value set to unit) measured for a capillary fiber filled with an agarose gel prepared using water as solvent (2% agar concentration). Since agarose gels display lower refractive index values than silica [9], the spectrum exhibits an oscillatory pattern with alternating regions of high and low transmission, as expected for antiresonant guidance within the filled capillary. In Fig. 2a, a baseline is calculated (dashed blue line) to allow for renormalizing the transmission spectrum for better visualization of its oscillatory behavior. Thus, by subtracting the baseline from the transmission spectrum, we obtain the plot shown in Fig. 2b, where the transmission maxima and minima can be readily observed.

The spectrum normalization procedure described above has been used to obtain the spectra shown in Fig. 3, which corresponds to capillary fibers filled with agar gels prepared using water-glycerol solutions with different concentrations.

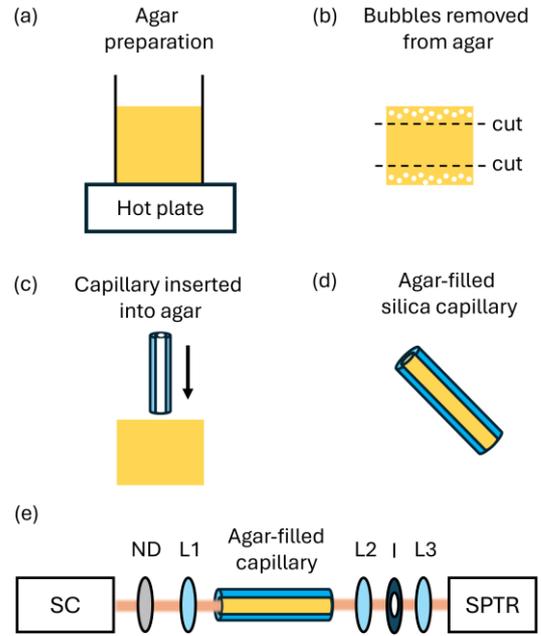

Fig. 1. (a) Preparation of the agar solution. (b) Removal of gel sections exhibiting bubble formation. (c) Insertion of an empty silica capillary into the agar gel. (d) Representation of the agar-filled silica capillary fiber. (e) Representation of the experimental setup for measuring the agar-capillary fiber transmission spectrum. SC: supercontinuum source; ND: neutral density filter; L1, L2, and L3: lenses; I: iris; SPTR: spectrometer.

As we will demonstrate in the following, the analysis of the spectral positions of the transmission peaks and dips forms the basis for determining the dispersion of the agar gels.

While one could directly employ Eq. (1) for estimating the refractive index of the agar gels (given that the capillary wall thickness and the refractive index of silica are known), this approach would involve assuming an initial value for $n_{agar}$ to determine the order $m$ corresponding to a specific transmission minimum occurring at $\lambda_m$ [12]. Alternatively, when the transmission spectrum is expressed in terms of the optical frequency, the free spectral range (FSR), $\Delta\nu$, remains approximately constant (thus independent of $m$), being described by Eq. (2), where $c$ is the speed of light [8]. This approach is more advantageous as it enables the direct calculation of an initial value for $n_{agar}$ by extracting $\Delta\nu$ from the measured spectra, thus eliminating the need for prior assumptions regarding the value of $n_{agar}$.

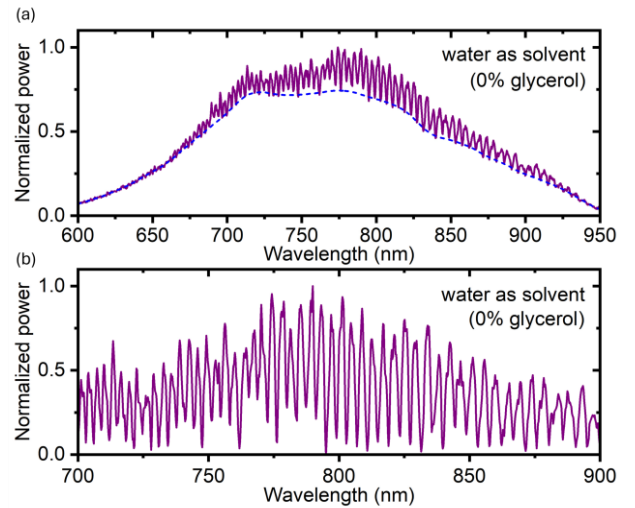

Fig. 2. (a) Normalized transmission spectrum of the agar-filled silica capillary fiber using water as solvent. The dashed blue line stands for the spectrum baseline. (b) Zoom in the 700 to 900 nm spectral region for better visualization of the data. The spectrum baseline has been used as reference for data renormalization.

$$\Delta\nu = \frac{c}{2t}\sqrt{n_{tube}^2 - n_{agar}^2} \qquad (2)$$

We thus conduct the following procedure for determining $n_{agar}$: initially, the measured transmission spectra are replotted in terms of the optical frequency. Subsequently, two consecutive maxima are selected from the central region of the spectrum (around 400 THz, approximately corresponding to a wavelength of 750 nm), from which the corresponding $\Delta\nu$ is obtained. Knowledge of $\Delta\nu$ then permits the calculation of an initial value for $n_{agar}$ by using Eq. (2). By plugging in this initial $n_{agar}$ value in Eq. (1), the order of the corresponding resonance, $m$, in the wavelength spectra can be determined. The orders of the other minima are then straightforwardly derived from this initial $m$ value (by adding or subtracting 1 from this initial value repetitively). This analysis yields the plots presented in Fig. 4, where the spectral positions of the minima are plotted as a function of their resonance order for capillaries filled with agar gels prepared using varying glycerol concentrations. As expected, due to the distinct refractive indices of agar gels with different glycerol content, resonances spanning different order ranges are observed within the considered wavelength interval.

Information presented in Fig. 4 allows for obtaining the dispersion of agar within the analyzed spectral range by using Eq. (1). The corresponding results are shown in Fig. 5, where the refractive index of the agar gels prepared using different glycerol concentrations has been plotted as a function of the wavelength. These results enable us to describe the dispersion behavior of the agar gels by fitting the data in Fig. 5 with a first-order Sellmeier equation according to Eq. (3), where $B$ and $C$ are fitting constants. The fitted curves are shown as dashed lines in Fig. 5 and the values of the fitted constants are summarized in Table 1. The Pearson $r^2$ values corresponding to the fits have been found to consistently exceed 0.98748.

$$n_{agar} = \sqrt{1 + \frac{B\lambda^2}{\lambda^2 - C}} \qquad (3)$$

Finally, to obtain a direct comparison with previously reported research [13], we used the fitted Sellmeier curves to determine the refractive index of the agar gels at 589 nm. For the 0% glycerol concentration, our derived refractive index is (1.3404 ± 0.0003), which shows a variation of 0.0064 compared to the literature value of 1.3340, corresponding to a percentage error of approximately 0.5%. At 20% glycerol concentration, the agar gel refractive index at 589 nm has been calculated as (1.3595 ± 0.0006), which is in close agreement with the reported 1.3615 (variation of about 0.2%). For the 40% glycerol concentration, our measurement yields (1.3851 ± 0.0003), differing by 0.0042 (approximately 0.3% error) from the literature value of 1.3893. At 60% glycerol concentration, our derived value has been found to be (1.3993 ± 0.0007), showing a variation of about 1.3% from the reported 1.4176 value. These comparisons demonstrate a good agreement between our experimental results and existing literature, particularly for lower glycerol concentrations, hence validating the refractometric approach reported herein. Variations in the water content due to agar syneresis and temperature deviations may explain the subtle deviations in the refractive index values.

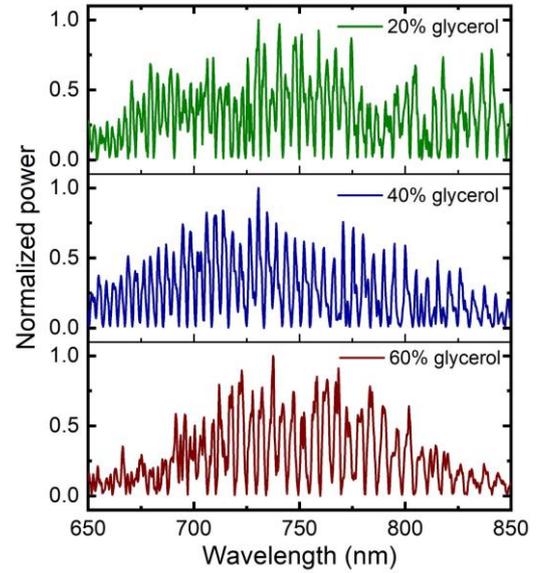

Fig. 3. Normalized transmission spectra (baseline subtracted) of the agar-filled silica capillary fiber using water-glycerol solutions with different concentrations.

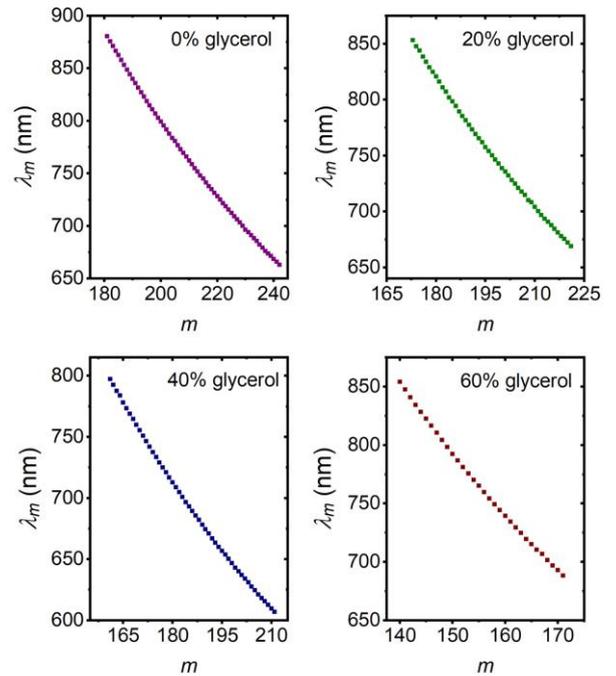

Fig. 4. Transmission minima ($\lambda_m$) as a function of the minima order ($m$) accounted from the measured transmission spectra of the agar-filled silica capillary fibers with different glycerol concentrations in the agar solvent.

In conclusion, we demonstrated a straightforward method for refractometric measurements based on the transmission characteristics of filled antiresonant capillary fibers. This approach offers advantages over conventional techniques, being compact, simple to implement, and capable of yielding dispersion data without requiring prior assumptions on the sample's refractive index. Here, we applied this method to characterize agar gels prepared with water-glycerol solutions of varying concentrations. Our results demonstrated good agreement with the literature, validating the reliability of the proposed technique. We understand that our work provides a relevant contribution to the development of practical tools for optical characterization, opening promising possibilities for refractometric analyses of a wide range of materials, and for the development of new refractive index sensing platforms.

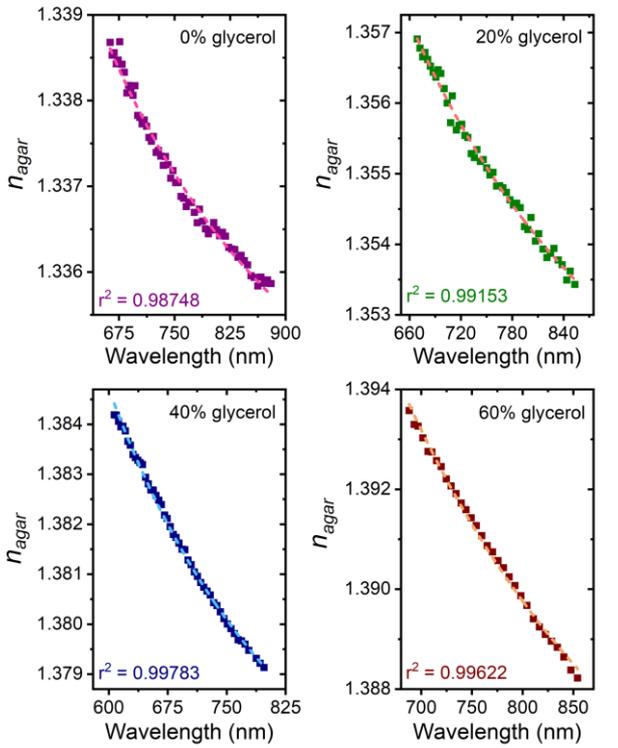

Fig. 5. Refractive index of the agar gels ($n_{agar}$) with different glycerol concentrations as a function of the wavelength. The dashed lines stand for the fitted curves using Eq. (3).

**Table 1. First-order Sellmeier constants for the agarose gels with different glycerol concentrations.**

| Glycerol concentration | $B$ | $C$ (μm²) |
|---|---|---|
| 0% | 0.7743 ± 0.0002 | 0.0097 ± 0.0001 |
| 20% | 0.8176 ± 0.0003 | 0.0126 ± 0.0002 |
| 40% | 0.8828 ± 0.0001 | 0.0136 ± 0.0001 |
| 60% | 0.9016 ± 0.0004 | 0.0205 ± 0.0002 |

**Acknowledgements.** The authors thank the support from the National Institute of Photonics (INFO), Brazil.

**Funding.** This work has been funded by FAPEMIG (RED-00046-23, APQ-00197-24, APQ-01401-22, APQ-01618-25), CNPq (309989/2021-3, 305024/2023-0, 402723/2024-4, 409174/2024-6), and FAPESP (2023/09181-0, 2024/02995-4, 2024/00998-6).